# The role of grain boundary scattering in reducing the thermal conductivity of polycrystalline $X$NiSn ($X$ = Hf, Zr, Ti) half-Heusler alloys

*Matthias Schrade[1], Kristian Berland[1], Simen N. H. Eliassen[1,2], Matylda N. Guzik[1,3], Cristina Echevarria-Bonet[3], Magnus H. Sørby[3], Petra Jenuš[4], Bjørn C. Hauback[3], Raluca Tofan[1], Anette E. Gunnæs[1], Clas Persson[1], Ole Martin Løvvik[1,5], and Terje G. Finstad[1]*

[1] Centre for Materials Science and Nanotechnology, Department of Physics, University of Oslo, Gaustadalléen 21, NO-0349 Oslo, Norway

[2] Department of Materials Science and Engineering, Norwegian University of Science and Technology, Norway

[3] Physics Department, Institute for Energy Technology, NO-2007 Kjeller, Norway

[4] Jožef Stefan Institute, Department for Nanostructured Materials, Ljubljana, Slovenia

[5] SINTEF Materials and Chemistry, Forskningsveien 1, NO-0314 Oslo, Norway

## *Abstract*

Thermoelectric application of half-Heusler compounds suffers from their fairly high thermal conductivities. Insight into how effective various scattering mechanisms are in reducing the thermal conductivity of fabricated $X$NiSn compounds ($X$ = Hf, Zr, Ti, and mixtures thereof) is therefore crucial. Here, we show that such insight can be obtained through a concerted theory-experiment comparison of how the lattice thermal conductivity $\kappa_{Lat}(T)$ depends on temperature and crystallite size. Comparing theory and experiment for a range of $Hf_{0.5}Zr_{0.5}NiSn$ and ZrNiSn samples reported in the literature and in the present paper revealed that grain boundary scattering plays the most important role in bringing down $\kappa_{Lat}$, in particular so for unmixed compounds. Our concerted analysis approach was corroborated by a good qualitative agreement between the measured and calculated $\kappa_{Lat}$ of polycrystalline samples, where the experimental average crystallite size was used as an input parameter for the calculations. The calculations were based on the Boltzmann transport equation and *ab initio* density functional theory. Our analysis explains the significant variation of reported $\kappa_{Lat}$ of nominally identical $X$NiSn samples, and is expected to provide valuable insights into the dominant scattering mechanisms even for other materials.

## *1. Introduction*

Waste heat harvesting by thermoelectric generators has great potential to increase the energy efficiency of many industrial processes. Recently, there has been a surge in efforts to improve the conversion efficiency of half-Heusler compounds, which are attractive thermoelectric materials for heat-to-electricity conversion in the medium to high temperature range from 400 to 800°C.

The promising thermoelectric properties of half-Heusler compounds $X$NiSn and $X$CoSb with $X$ = Ti, Zr, or Hf in combination with the abundance of constituent elements, chemical stability, and non-toxicity, have made them one of the most studied material systems for waste-heat harvesting applications [1-3]. The thermoelectric performance of these half-Heusler materials is, however, limited by their relatively high lattice thermal conductivity $\kappa_{Lat}$ [4]. Aiming to scatter phonons at different length scales, several strategies to reduce $\kappa_{Lat}$ have been pursued including atomic substitution to increase mass disorder [5-8], nanostructuring by mechanical processing [9, 10] and segregation of nanoinclusions [11-13].

The highest figure of merit, $zT$, of $X$NiSn half-Heusler alloys so far reported is around 1.5 at 700 K for Sb-doped $Hf_{0.25}Zr_{0.25}Ti_{0.5}NiSn$ [14, 15], and has been attributed to the unusually low thermal conductivity of those particular samples. Insight into the contribution of different scattering processes can help to understand such results better and, in turn, assess the potential of the various strategies to reduce $\kappa$.

In recent years, there has been much progress in predicting thermal transport properties from *ab initio* calculations. In part, this is due to increased computational power, but even more so it is due to the implementation of effective algorithms and toolboxes [16, 17]. Such methods can be used to optimize the thermal conductivity by calculating the effect of different scattering mechanisms.

While theoretical transport calculations can give insight into the effect of different scattering mechanisms, agreement between calculated and measured values of $\kappa$ does not directly reveal which scattering mechanisms dominate in a given fabricated sample, as many different scattering mechanisms can contribute to reducing the thermal conductivity.

Here, we use a concerted theory-experiment comparison to show that grain-boundary scattering or similar mechanisms targeting the low frequency phonons can explain the significant spread in measured $\kappa_{Lat}$ for nominally identical samples.

In the first step, we fabricated samples of different compositions, analyzed them experimentally and compared the results with those of density functional theory (DFT) calculations. The fabrication procedure led to samples with average crystallite sizes of smaller than 100 nm, as deduced by synchrotron radiation X-ray measurements. Using the measured values as grain size parameter in the theory, we find, without tuning any other parameters, that that DFT-based calculations agree quite well with the experimental $\kappa_{Lat}$ for several samples. This comparison demonstrates the importance of grain-boundary scattering in our samples for many different compositions.

In the second step, we analyzed the significant spread in measured $\kappa_{Lat}$ for a range of ZrNiSn and $Hf_{0.5}Zr_{0.5}NiSn$ samples reported in literature. For these samples we did not have access to the average crystallite size. However, we could still compare the role of point defect and grain boundary scattering, by comparing both the magnitude and temperature dependence of $\kappa_{Lat}$.with corresponding theoretical predictions. This led to far better agreement, when grain-boundary scattering rather than localized point defect scattering was assumed to be the main mechanism causing the spread in $\kappa_{Lat.}$

## *2. Methods*
### *2.1. Experimental methods*
Polycrystalline samples of the different compositions were obtained and characterized as described in the following. Stoichiometric amounts of Ti (Alfa Aesar, 99.5%), Zr, (GoodFellow 99.8%), Hf (Alfa Aesar, 99.7%), Ni (GoodFellow 99.99%), and Sn (GoodFellow, 99.995%) were first arc-melted under Ar atmosphere. The samples were turned and re-melted several times to ensure homogenization. The resulting pellets were manually ground and sealed in evacuated quartz ampules. The samples were then annealed at 1123 K for seven days and subsequently quenched in ice water. In turn, fine-grained powders were obtained by extended ball milling using a planetary mill (Fritsch P7) at a rotational velocity of 450 rpm under Ar atmosphere at room temperature. The powders were thereafter densified using a spark plasma sintering (SPS) machine (Dr. Sinter 511S/515S). Prior to the sintering, the powders were transferred into a 15 mm diameter graphite die and sintered in vacuum at 1123 K for 10 min. The applied pressure varied from 65-85 MPa, depending on sample composition. The surfaces of the sintered pellets were ground with SiC paper to remove possible surface contamination. All sintered samples had a volumetric density of at least 93 % of their theoretical value, Table S.1.

The nominal composition of the investigated samples can be marked in a pseudo-ternary phase diagram of (Hf,Zr,Ti)NiSn, as shown in Fig. 1. The marker symbols and colors identify the same compositions throughout the paper. We will refer to the corners of the diagram in Fig. 1 as the "unmixed", and the others as "mixed" compounds.

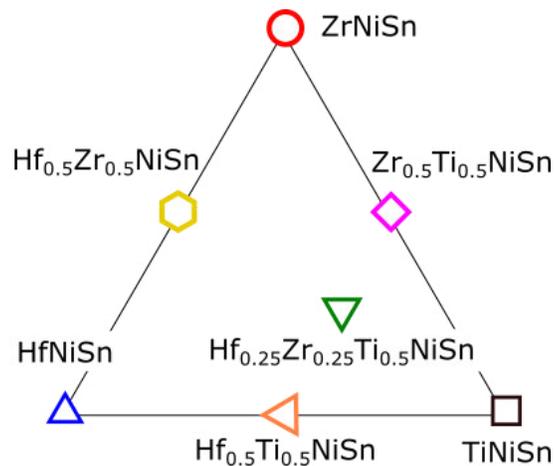

**Figure 1.** Pseudo-ternary phase diagram for (Hf,Zr,Ti)NiSn. The marker symbol and color identify the investigated nominal compositions, as explained in the text.

Prior to characterization, the pellets were cut into smaller pieces with a diamond saw. A piece of each composition was crushed in a mortar for structural analysis. Synchrotron radiation powder X-ray diffraction (SR-PXD) data were collected at room temperature at the Swiss-Norwegian beamline at the European Synchrotron Radiation Facility (ESRF, Grenoble, France, station BM01, $\lambda$ = 0.6973 Å). In all SR-PXD experiments powders were sealed in boron-glass capillaries of 0.3 or 0.5 mm in diameter. Diffraction data were collected with a Pilatus2M 2D area detector, integrated and analyzed by conventional Rietveld refinement through the FullProf Suite [18]. The average crystallite size, $d$, and an isotropic lattice strain, $\eta$, were extracted from diffraction line broadening. The diffraction profiles were modeled using a Thomson-Cox-Hastings pseudo-Voigt function to simulate the peak shapes of half-Heusler phases. The instrumental resolution function was obtained from the refinement of $LaB_6$.

Microstructural characterization was performed by scanning electron microscopy (SEM) and transmission electron microscopy (TEM). SEM characterization of polished sample surfaces was carried out using a Hitachi tabletop TM3000 scanning electron microscope (SEM) with back-scatter electron detector and a Quantax70 detector for X-ray energy dispersive spectroscopy (EDS). For the TEM analysis, a JEOL2100F instrument with an acceleration voltage of 200 kV was used. A piece of the sintered sample was crushed in ethanol and the mixture was left to dry on a Cu-grid covered with lacey C.

To measure the thermal conductivity $\kappa$, thermal diffusivity measurements were first performed on the same sample under nitrogen flow in a laser flash apparatus (Netzsch LFA 451). $\kappa$ was obtained from the thermal diffusivity $D$ by $\kappa = D\rho c_P$ using the geometrical mass density $\rho$ and the specific heat capacity $c_p (\approx c_V)$ as calculated by DFT, Figure S.3. The in-plane electric conductivity was measured under argon atmosphere using a custom-made set-up detailed elsewhere [19].

## 2.2. Theoretical methods

The lattice thermal conductivity was calculated using the Boltzmann transport equation in the relaxation-time approximation, relying on inter-atomic forces obtained with DFT. For the DFT part, we used the VASP [20] software package with the Perdew-Burke-Ernzerhof (PBE) exchange-correlation functional [21]. The PHONO3PY [17] package was used to obtain the phonon dispersion and relaxation times, and in turn the lattice thermal conductivity.

Three phonon scattering mechanisms [22] were considered: i) three-phonon scattering arising from the anharmonicity of the inter-atomic potential, ii) mass-disorder scattering [23], with contributions both from the natural variation of isotopes and, for mixed compounds, from mass contrast on the $X$ site, and iii) grain boundary scattering. The latter was accounted for with a simple diffusive scattering model with a scattering rate of the form $v_{iq}/d$ [24], where $q$ is the phonon wavevector and $i$ is the band index. The total relaxation time for each $q$ and band $i$ was obtained using Matthiessen's rule. To model mixed compounds, we used the virtual crystal approximation, obtaining effective alloy properties by making a linear average of the atomic masses and inter-atomic forces obtained for HfNiSn, ZrNiSn, and TiNiSn. This approach and other computational choices are detailed in a preceding paper [25]. This paper highlights the importance of accounting for the changing nature of the vibrational modes with composition. In section 5, we also generate results for ZrNiSn and $Hf_{0.5}Ti_{0.5}NiSn$ including a generic point-defect scattering with an adjustable scattering strength. Given that the point defect is left unspecified, it is simply emulated as a mass-disorder scattering with the same mass variance on the $X$, Ni and Sn site. This allows us to compare how these scattering mechanisms affect the temperature dependence of $\kappa_{Lat}$.

## 3. Results

### 3.1 Theoretical results

Fig. 2 shows $\kappa_{Lat}$ of the pseudo-ternary (Hf,Zr,Ti)NiSn as calculated by DFT at 400 K, assuming a constant average grain size of 100 nm. The lattice thermal conductivity of the three unmixed half-Heusler compounds is found to be around 7 WK$^{-1}$m$^{-1}$, but decreasing steeply with increasing atomic substitution. The lowest $\kappa_{Lat}$ is predicted along the Hf$_{1-x}$Ti$_x$ line, which has the largest mass contrast, resulting in values as low as 1.7 WK$^{-1}$m$^{-1}$. A similar behavior can be observed for Hf$_{1-x}$Zr$_x$NiSn compositions, with values down to 2.4 WK$^{-1}$m$^{-1}$. However, for the Zr$_{1-x}$Ti$_x$NiSn compositions, the minimum is around 4 WK$^{-1}$m$^{-1}$, much larger than for the Hf containing compositions. The result can be explained by an interplay between the varying mass contrast and the shifting nature of the phonon modes [25]. In short: the scattering is strongest, if the atomic site contributing the most to the phonon density of states is also the site with largest mass contrast.

The calculations are performed for homogeneous materials, while the experimentally investigated, mixed samples show a varying degree of separation into phases of different composition. Nonetheless, the comparison with experiment can be rationalized by effective medium considerations, i.e., $\kappa_{Lat,Mix} = \sum_i c_i \times \kappa_{Lat,i}$, where $c_i$ is the relative concentration of phase $i$, and $\kappa_{Lat,i}$ is its lattice thermal conductivity [26]. $\kappa_{Lat,i}$ is only weakly dependent on composition for mixed compositions away from the corners of Fig. 2. Therefore, $\kappa_{Lat,i}$ is rather similar for the different phases present in each of the samples investigated here, so that $\kappa_{Lat,Mix}$ can be treated as representative for a homogeneous, i.e. not phase separated, sample of average chemical composition.

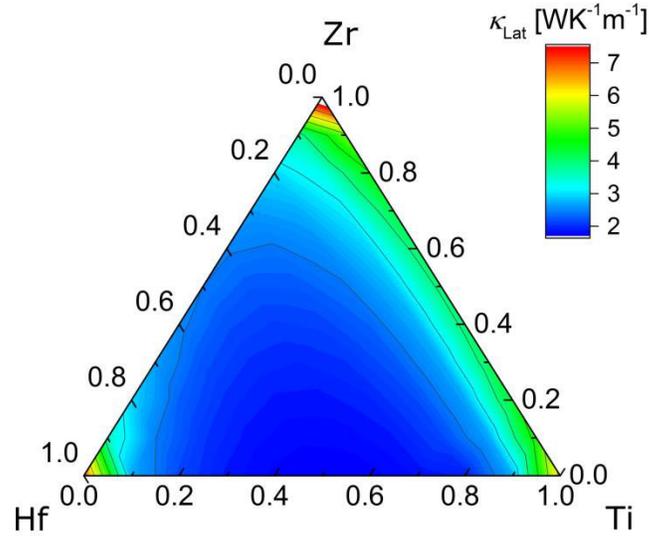

**Figure 2.** $\kappa_{Lat}$ obtained by DFT in the pseudo-ternary phase diagram of (Hf,Zr,Ti)NiSn. Values are calculated at 400 K and using an average grain size of 100 nm. Unmixed compositions are predicted to have a significantly higher $\kappa_{Lat}$ than the rest.

### *3.2. Structural characterization*

Main reflections of collected diffraction patterns were indexed with half-Heusler phases of different composition. In addition, samples exhibited small amounts of Sn and full-Heusler phases. Rietveld refinement revealed that the mixed samples crystallized as different half-Heusler phases, with a mixed atomic occupancy on the 4*a* crystallographic site, in qualitative agreement with earlier reports [27]. The composition of the individual half-Heusler phases was calculated from Vegard's law, based on the lattice parameter values refined for the unmixed *X*NiSn [28]. As an example, Fig. 3 (a) shows a SR-PXD pattern of $Ti_{0.5}Hf_{0.5}NiSn$ with an apparent splitting of the Bragg reflections and the corresponding refined profile. From the Rietveld analysis of the diffraction profiles, the average values of crystallite sizes of the half-Heusler phases were obtained. The refined average crystallite sizes range from 60 to 100 nm, as qualitatively confirmed by TEM (Fig. 3 (b)) and similar to previously reported values for half Heusler samples prepared in a similar way [29, 30]. An overview of the refinement of all investigated samples is given in the supporting information, S.1, and is complemented by additional TEM pictures, illustrating the distribution of crystallite sizes within the sample, Fig. S.1.

These results were corroborated with microstructural analysis using SEM. In agreement with the results from SR-PXD, the SEM analysis also shows phase separation into several half-Heusler phases, as illustrated by the micrograph shown in Fig. S.2. The chemical compositions, as obtained by EDS, agree well with those obtained from the SR-PXD data.

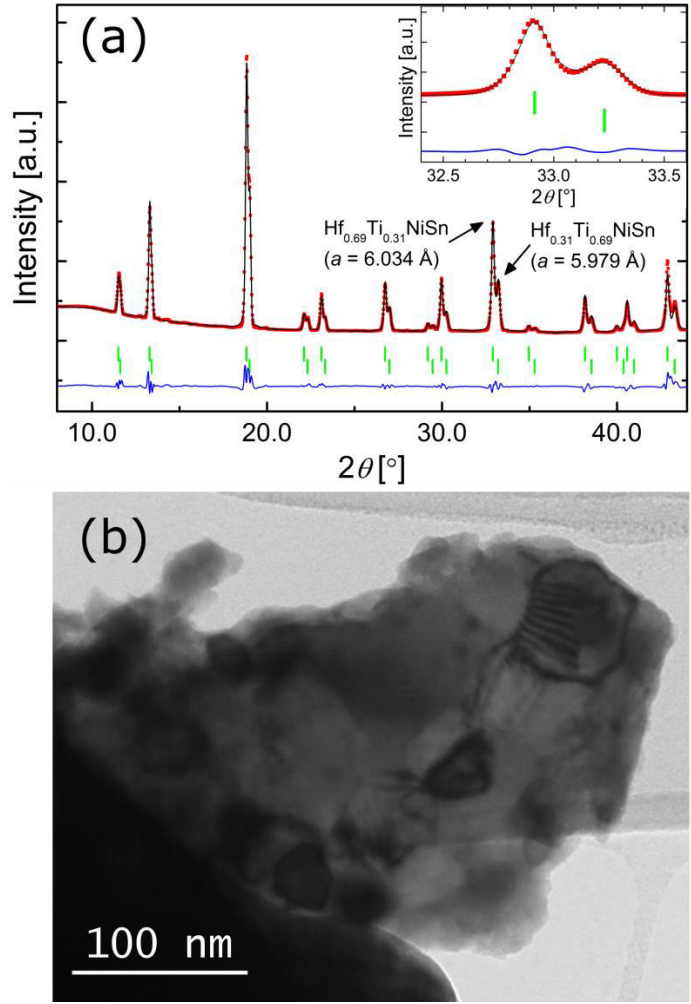

**Figure. 3.** (a) Observed (red), calculated (black) and difference (blue) SR-PXD patterns for $Ti_{0.5}Hf_{0.5}NiSn$, with a clearly visible splitting of the Bragg reflections ($R_p$ = 5.37, $R_{wp}$ = 8.25). Vertical bars indicate Bragg peaks positions of contributing phases. (b) TEM micrograph of ZrNiSn after sintering, showing several nano-sized crystallites.

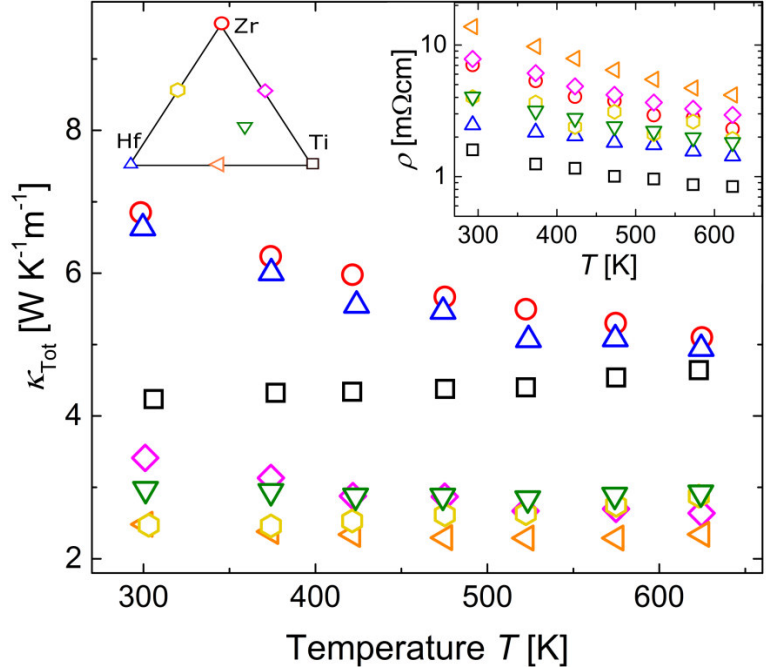

**Figure 4.** The experimental total thermal conductivity $\kappa_{Tot}$ versus temperature $T$. $\kappa_{Tot}$ shows a weak temperature dependency. The mixed samples exhibit a lower $\kappa_{Tot}$ than the unmixed samples. The inset shows the electrical resistivity $\rho$ against temperature $T$.

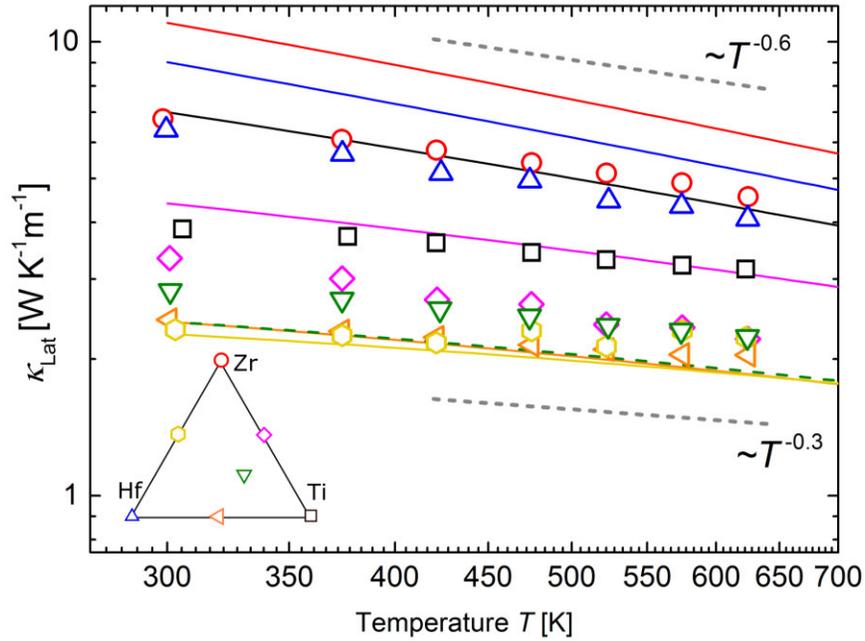

**Figure 5.** Lattice-part of the thermal conductivity $\kappa_{Lat}$ against temperature $T$. Lines represent the DFT results, using the same colors as the experimental symbols to identify composition. Alloy- and grain boundary scattering processes are included,

and the average crystallite size as obtained from SR-PXD is used as experimental input to the DFT calculations.

### *3.3. Experimental thermal properties*

The total thermal conductivity, $\kappa_{Tot}$, of the investigated samples is shown as a function of temperature $T$ in Fig. 4. The values of $\kappa_{Tot}$ for the mixed samples are almost independent of temperature, and quite similar for the different mixed compositions. The unmixed samples with $X$ = Hf and Zr exhibit a significantly higher $\kappa_{Tot}$ and a more pronounced temperature dependence, with values around 7 WK$^{-1}$m$^{-1}$ at 300 K, reducing to 5 WK$^{-1}$m$^{-1}$ at 650 K. In contrast, $\kappa_{Tot}$ of the TiNiSn sample has both a lower magnitude, around 4 WK$^{-1}$m$^{-1}$, and $\kappa_{Tot}$ depends far less on temperature. Thus, the result for TiNiSn falls in between the corresponding values for the other unmixed and the mixed samples. To estimate the electronic contribution to the thermal conductivity, the electrical resistivity $\rho$ was measured for different temperatures $T$, with results shown in the inset of Fig. 4. $\rho$ decreases with increasing temperature, with room-temperature values ranging from 2 to 14 mΩcm, which is in general agreement with values reported in the literature for similar compositions [28, 31, 32]. We estimated the electronic contribution to the thermal conductivity $\kappa_{El}$ using the Wiedemann-Franz law:

$$\kappa_{El} = L \cdot T \cdot \rho^{-1}$$

(2)

$L$ is the Lorenz function, here taken as constant [33]. Previous studies have found $L$ to be lower than the Sommerfeld limit of $2.44 \times 10^{-8} W\Omega K^{-2}$ in $X$NiSn compounds without intentional doping [12, 34]. We use the value of $2 \times 10^{-8} W\Omega K^{-2}$ to calculate $\kappa_{El}$ for all samples. Our analysis is not particularly sensitive to this choice, since $\kappa_{El}$ is only a minor contribution to $\kappa_{Tot}$ for all samples considered here. The lattice part of the thermal conductivity, $\kappa_{Lat} = \kappa_{Tot} - \kappa_{El}$, is shown in Fig. 5. The values of $\kappa_{Lat}$ decrease slightly with increasing temperature for all samples. In the analysis in section 4, we will find it helpful to describe the temperature dependency of $\kappa_{Lat}$ in terms of a power law, i.e. $\kappa_{Lat}(T) \sim T^{-x}$. For $X$ = Hf and Zr, we find the exponent $x$ to be approximately 0.6. $x$ is lower for the mixed samples, e.g., approximately 0.3 for $X$ = Hf$_{0.5}$Ti$_{0.5}$.

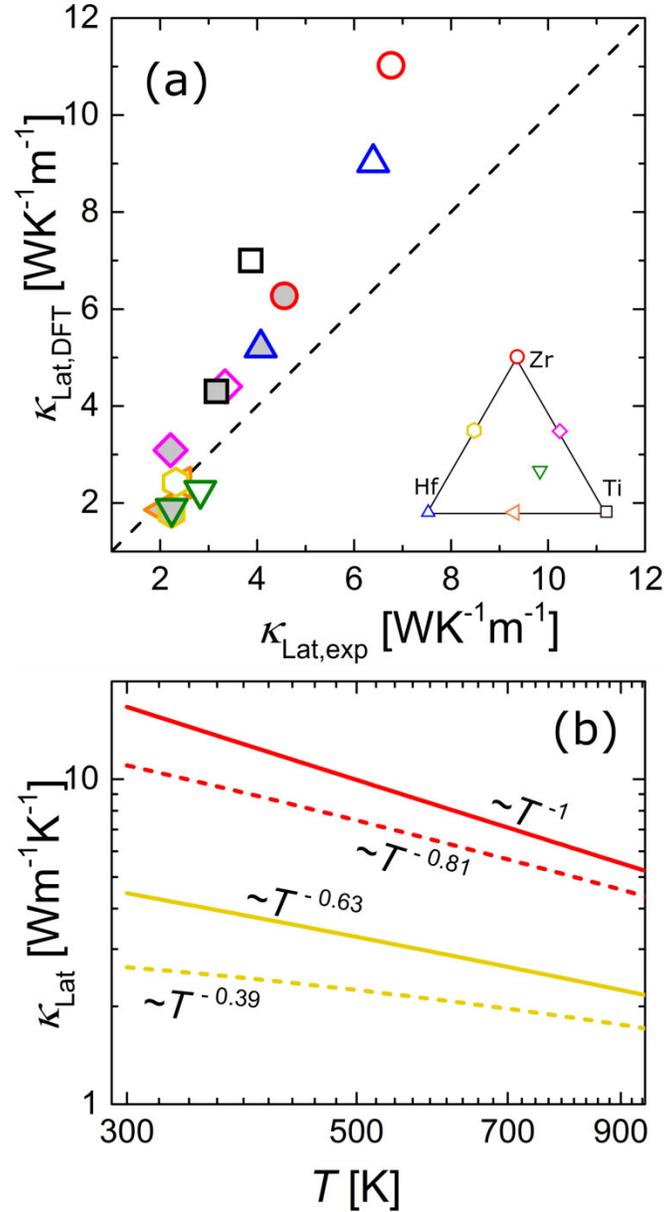

**Figure 6**. (a) Comparison of the experimental $\kappa_{Lat,exp}$ with the DFT-based results $\kappa_{Lat,DFT}$. Open and grey-filled symbols represent 300 K and 625 K, respectively. The dashed line represents the ideal 1:1 correspondence between theory and experiment.

For low values of $\kappa_{Lat}$, DFT and experiment agree well, while theory increasingly overestimates $\kappa_{Lat}$ for increasing $\kappa_{Lat,exp}$. **(b)** Temperature dependence of $\kappa_{Lat,DFT}$ for $X$ = Zr (red) and Hf$_{0.5}$Zr$_{0.5}$ (yellow). Solid lines are calculated without grain boundary scattering, while dashed lines assume an average crystallite size of 70 nm. Both grain boundary and point defect scattering reduce the absolute value and the temperature dependence of $\kappa_{Lat}$.

### *3.4. Comparison between experiment and DFT*

For each of the samples, we use the average crystallite size as obtained by analysis of the SR-PXD data as input parameter for the calculations, specifying the grain size parameter (Section S.1., supporting information). The calculated $\kappa_{Lat,DFT}$, which includes phonon scattering resulting from anharmonic, mass disorder, and grain boundary scattering as a function of temperature is given by the solid curves in Fig. 5.

The overall agreement between experimental and theoretical results is good. For the mixed samples, the theory agrees well with the experimental data, while it overestimates $\kappa_{Lat}$ for the unmixed compounds (Fig. 6 (a)). Further, the temperature dependence of the experimental $\kappa_{Lat}$ is well reproduced by DFT for all samples. The overestimation of $\kappa_{Lat}$ for the unmixed samples can be qualitatively justified by the presence of additional point defects or impurity phases, such as antisite defects between the *X* and Sn sublattice, which have been a source of some controversy in the literature [35-37]. Another point defect, not captured explicitly in the DFT-based calculations presented so far, is Ni interstitials. These are particularly prominent for TiNiSn [12, 38-41], but have also been reported for ZrNiSn [42, 43]. For the mixed samples, the significant disorder on the *X* sublattice constitutes an effective scattering mechanism for the more energetic phonons. Thus additional point defects or impurities are less likely to affect $\kappa_{Lat}$ significantly, unless their density is enormous. In order to systematically evaluate the quantitative effect of point defects like antisites and interstitials on $\kappa_{Lat}$ in DFT-based calculations, software tools must be further refined, even if much progress has been made recently in this direction [44]. The residual discrepancy between theory and experiment could be due to the simple account of grain-boundary scattering and the mean crystallite size can only be viewed as a good estimate for an appropriate grain-boundary scattering parameter. We also note that our DFT calculations do not include thermal expansion which could also affect the heat capacity and lattice thermal conductivity somewhat. However, since the thermal expansion coefficient of approximately $10^{-5}$ $K^{-1}$ of these compounds [45] is quite modest compared to other thermoelectric materials, we do not expect a significant impact on the obtained $\kappa_{Lat}$, in particular with respect to the general accuracy of the results. Furthermore, to properly match such an analysis with experiment, more detailed microscopic structural information is required, namely a refinement of the individual occupancies of all sublattices, including the vacancy site, not to mention the potential presence of more complicated defects.

## *4. Analysis*

Grain boundary and point defect scattering have two distinct effects: they both reduce the absolute value of $\kappa_{Lat}$ and its temperature dependence, as illustrated in Fig. 6 (b). Since point defects predominantly target the more energetic phonons, whereas grain boundary scattering more strongly targets the fast low-energetic acoustic phonons [25], increasing the scattering strength of one of these mechanisms affects the temperature dependence somewhat differently.

We will in the following use this difference as an analysis tool to explore whether the experimental samples primarily scatter short- or long-range phonons, thus giving an indication of the key mechanisms reducing the lattice thermal conductivity.

To do so, we collected several $\kappa_{Lat}(T)$ data sets for ZrNiSn and Hf$_{0.5}$Zr$_{0.5}$NiSn from the literature. These compositions were chosen as representative examples of an unmixed and mixed composition, with sufficient reported data available. Both experimental and theoretical data can empirically be described as a power law, $\kappa_{Lat} \propto T^{-x}$ in the investigated temperature range from 300 to 700 K. The experimental data for our samples and the fitted curves of our samples are shown in Fig. S.4, (supporting information).

Fig. 7 shows the temperature exponent $x$ of the individual data sets against the room temperature value of $\kappa_{Lat}$. For both compositions, the reported values for $\kappa_{Lat}$ vary significantly and one can observe a stronger temperature dependency $x$ for those samples with a higher $\kappa_{Lat}(300K)$.

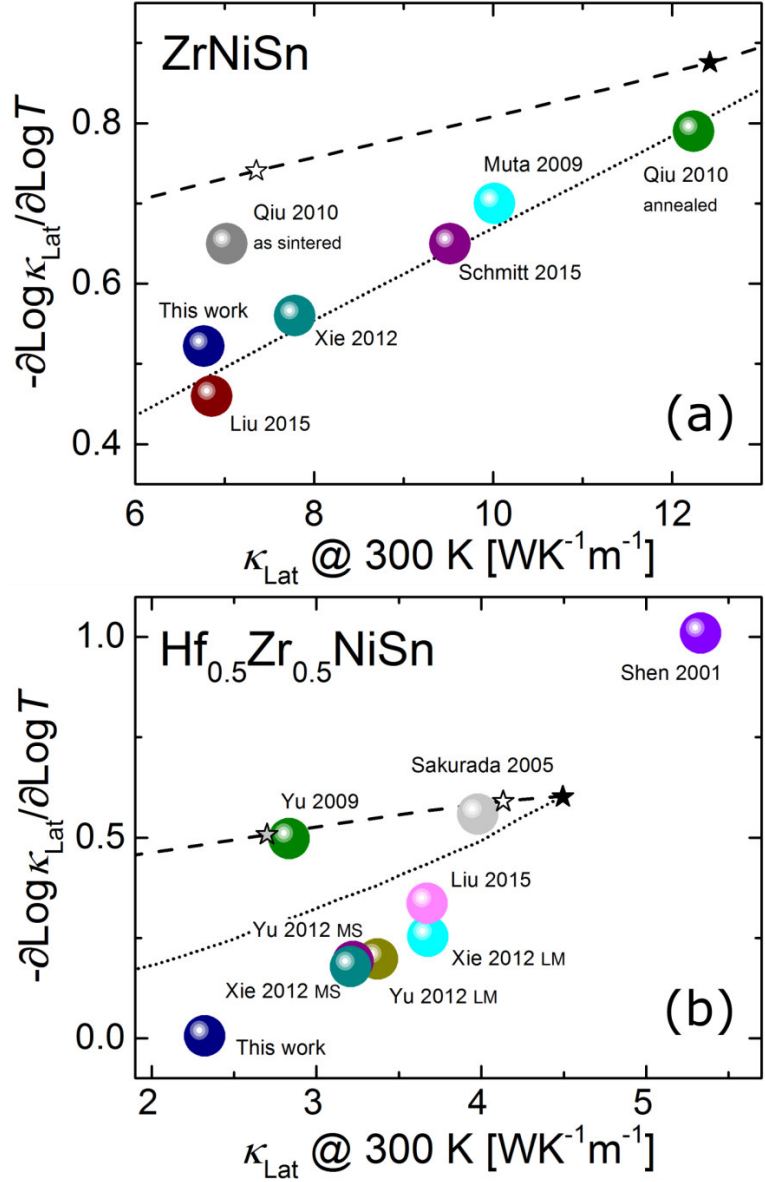

**Figure 7. (a)** Temperature exponent $x = -\partial \log \kappa_{Lat}/\partial \log T$ of reported data for ZrNiSn against the room temperature value of $\kappa_{Lat}$ [5, 30, 36, 42, 46]. The curves are calculated by DFT, using a point defect-like scattering term ($\propto \omega^4$, dashed curve) and a grain boundary scattering term ($\propto v_g/d$, dotted curve), respectively. Black and white stars indicate a scattering strength of $10^{-3}$ and $10^{-2}$, respectively. **(b)** Similar to (a), but for $Hf_{0.5}Zr_{0.5}NiSn$ [5, 8, 30, 47-49]. LM and MS refer to levitation melting and melt spinning. Black, white, and grey stars indicate a scattering strength of $10^{-3}$, $10^{-2}$, and $10^{-1}$, respectively. The correlation between $\kappa_{Lat}(300K)$ and $x$ in both (a) and (b) can best be described by different crystallite sizes of the investigated samples.

To explain the correlation between $\kappa_{Lat}$ and $x$ we examine three different scenarios; that it is primarily due to i) electron-phonon scattering, ii) point defect scattering, and iii) grain boundary scattering.

i) Electron-phonon scattering has been included in some earlier studies [5, 34] as a means to reproduce the experimental $\kappa_{Lat}$ within the Callaway model [50]. The relaxation time of electron-phonon scattering depends on material parameters like the deformation potential and the carrier effective mass - which should not vary between samples of nominally identical composition - and on the charge carrier concentration, which could be different for all investigated samples. If this was the dominant mechanism, one would expect a correlation between the temperature exponent $x$, or $\kappa_{Lat}$(300K), and the absolute value of the electrical conductivity, which is directly related to the charge carrier concentration. However, we find no such correlation in the reported data sets, Fig. 8, which indicates that electron-phonon scattering is unlikely to be a relevant process in describing the thermal conductivity in these samples.

This argument is further supported by the minute variation of $\kappa_{Lat}$ observed in previous doping studies [8, 30, 49]. There, the charge carrier concentration was varied for constant composition *X*, as evidenced by a systematic variation of Seebeck coefficient and electrical conductivity. However, no corresponding trend of the lattice thermal conductivity was reported for the same samples.

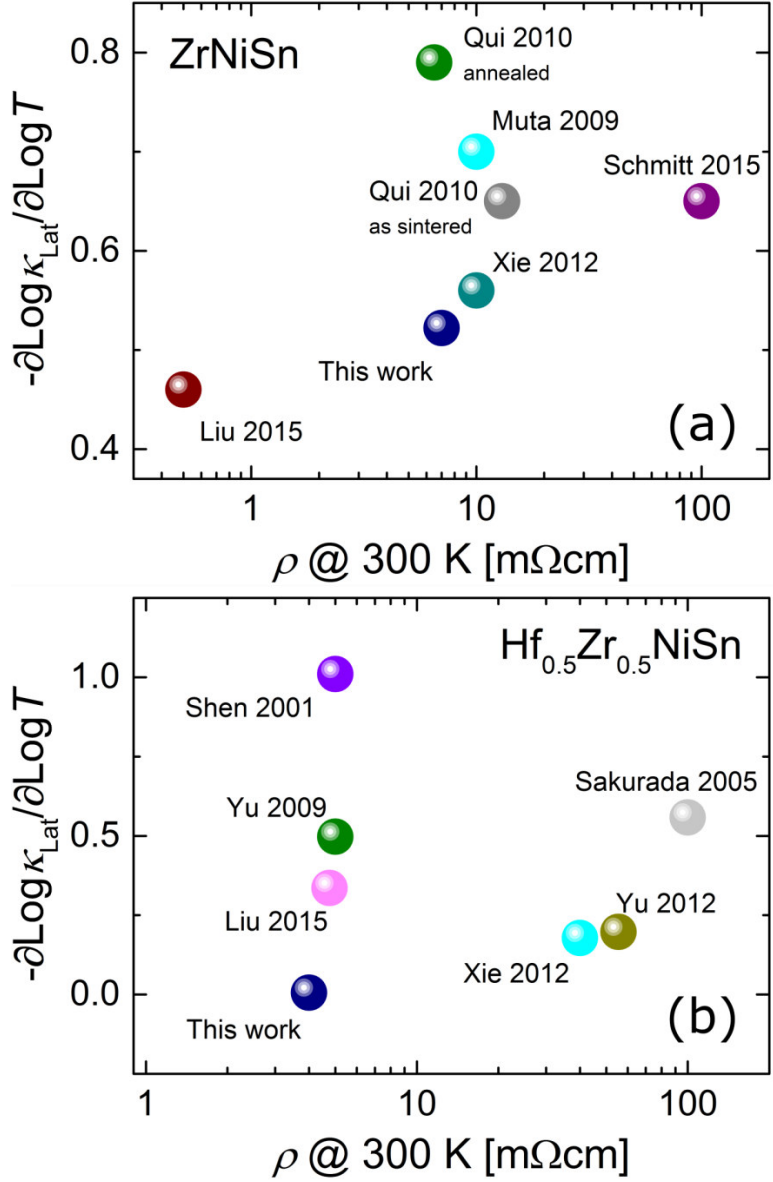

**Figure 8.** The temperature exponent $x$ against the electrical resistivity $\rho$ at room temperature for the samples shown in Fig. 7. Electron phonon scattering is expected to scale with the charge carrier concentration, which in turn scales $\rho$. No obvious correlation is, however, observed for both ZrNiSn **(a)** and $Hf_{0.5}Zr_{0.5}NiSn$ **(b)**.

ii) For ZrNiSn, the different values of $\kappa_{Lat}(300K)$ reported in the literature could be related to different concentrations of point defects in the individual samples. For example, using state-of-the-art computational methods, Katre *et al.* reproduced $\kappa_{Lat}(T)$ values reported by Qiu *et al.* [36] using different sets of Zr-Sn antisite and Ni/vacancy defects [44]. Their theoretical analysis indicated that in ZrNiSn, Ni/vacancy anti-site is a much stronger, and therefore more important, phonon scattering mechanism than Sn/Zr antisites. In our study, to mimic the effect of unintentional point defects in ZrNiSn, we artificially added a scattering term $\propto \omega^4$ into our DFT calculations. The result is included as a dashed line in Fig. 7. Point defects can qualitatively account for the observed correlation between $x$ and $\kappa_{Lat}$(300K), however, the quantitative agreement is less good. Using $\propto \omega^6$, which could be more appropriate for antisite defects [44] would yield an even poorer agreement, since this would make $\kappa_{Lat}$ vary even less as a function of temperature. We note that such simplified power-law form may not necessarily match the computed scattering rate in a more detailed description [44]. However, this would not greatly impact our argument, as point-defect scattering would still scatter high-frequency phonons much stronger than low-frequency phonons.

iii) Grain boundary scattering also reduces both the absolute value of $\kappa_{Lat}$ and $x$, as illustrated in Fig. 6(b). We therefore varied the parameter $d$ in the scattering term $\propto v_g/d$ in the Boltzmann transport calculations for ZrNiSn to account for a variation of the average crystallite size between the samples. The result (dotted line in Fig. 7) describes the reported data well, both qualitatively and quantitatively. This result suggests that the observed range of reported values for $\kappa_{Lat}$ in ZrNiSn could in large part be due to the different average crystallite sizes in the investigated samples, which is dependent on the specific synthesis and sintering process. This is not to say that point defects do not play any significant role, in particular so for the samples of Qui *et al.* [36] and Muta *et al.* [46], which are less temperature dependent than the others. Nonetheless, our analysis suggest that grain boundary scattering or a related mechanism is the key to explain the overall spread in measured $\kappa_{Lat.}$

We performed a similar analysis of $Hf_{0.5}Zr_{0.5}NiSn$. Just as in the case of ZrNiSn, the observed correlation between $x$ and $\kappa_{Lat}$ can be rationalized by a variation of the average crystallite size in the different samples, dotted line in Fig. 7 (b). Even if the dashed line shows a similar agreement as for the ZrNiSn case, it is much harder to rationalize the results in terms of additional point defects in the $Hf_{0.5}Zr_{0.5}NiSn$ case. This is because an enormous defect concentration would be required to reproduce the variation observed in the experimental data. To illustrate this point, the black and white star symbols in Figure 7 show the effect of introducing a point-defect scattering strength of $10^{-3}$ and $10^{-2}$. This range is enough to reproduce the range of measured $\kappa_{Lat}$ of ZrNiSn samples, so it gives an indication of a reasonable point-defect scattering strength variation in $Hf_{0.5}Zr_{0.5}NiSn$, if it was in fact point defects causing the variation in $\kappa_{Lat}$. However, this range of scattering strength barely changes $\kappa_{Lat}$, as indicated by the black and white star markers in Fig 7b). In fact, a ten times larger scattering strength is needed (gray star) to reasonably cover the range. But this corresponds to a total point-defect scattering strength considerably stronger than the total mass-disorder scattering in these samples, and structural integrity of the samples would be questionable at best. This comparison of theory and experiment thus clearly indicates that the spread in $\kappa_{Lat}$ must arise from a mechanism targeting the low-energetic phonons, such as grain boundary scattering.

We note that even if our results agree well with grain boundary scattering in the diffusive limit, other related mechanisms could also bring down $\kappa_{Lat}$ in a similar way. For instance, nanoinclusions and precipitates would also more strongly target low-frequency phonons and could therefore give similar agreement with experiment. The nature of boundary scattering could also be somewhat different in mixed and unmixed samples, as the latter tends to show phase separation, another mechanism to lower $\kappa_{Lat}$ proposed in the literature [15, 51, 52]. However, in practice, such mechanism could also be accounted for in a rough manner by the effective grain size *d*, parameterizing the strength of the scattering.

## *5. Summary*

We have investigated the role of grain-boundary scattering in lowering the lattice thermal conductivity of *X*NiSn (*X* = Hf, Zr, and Ti) half-Heusler alloys using a combination of experimental and computational techniques. First, several different compositions were fabricated using average crystallite sizes between 60-100 nm, as determined by SR-PXD. The thermal conductivity of these samples was as low as 2.5 WK$^{-1}$m$^{-1}$ at room temperature for both Hf$_{0.5}$Zr$_{0.5}$NiSn and Hf$_{0.5}$Ti$_{0.5}$NiSn compositions. An *ab initio* DFT-based analysis attributed this to a combination of mass disorder scattering arising from atomic substitution on the *X* sublattice and phonon scattering at grain boundaries.

The good agreement between DFT and experimental results further allowed us to discuss the origin of the significant spread in values of $\kappa_{\text{Lat}}(T)$ reported in the literature on nominally identical samples. We theoretically compared how point defects and grain-boundary scattering affected both the lattice thermal conductivity and its temperature dependence, parametrized in the form $T^{-x}$. Additional grain boundary scattering agreed better with the experimental trends than enhanced point defect scattering. This detailed comparison of measured properties and DFT calculations is an effective strategy to reveal the dominant scattering mechanisms of fabricated samples. The importance of grain-boundary in lowering the thermal conductivity suggests that nanostructuring is a promising strategy for minimizing the lattice thermal conductivity in half-Heusler and similar compounds. However, further studies of the electronic properties are needed, in order to assess the potential of nanostructuring for the overall thermoelectric performance of *X*NiSn half-Heusler materials.

## *Acknowledgement*

This work was funded by the Research Council of Norway within the THELMA project (No. 228854). The Notur consortium is acknowledged for computational resources.

## *Author contributions*

MS and TGF initiated the project of experimentally studying the thermal conductivity of *X*NiSn. Samples were made by CEB, MNG, and PJ. SR-PXD experiments were conducted and analyzed by CEB, MNG, and MHS, with BCH supervising data interpretation. SEM and TEM characterization was performed by RT and AEG, respectively. MS performed the thermoelectric transport measurements, supervised by TGF. Initial DFT-based calculations were done by SNHE, supervised by KB and OML, while KB performed further calculations to provide the presented analysis. CP supervised the computational efforts and secured funding for the whole project. MS and KB performed the combined theory-experiment analysis and wrote the manuscript. All authors have critically contributed to data and manuscript discussion and the process was administered by MS.

## *Additional information*

**Availabilty of materials and data:** Raw data and materials of the presented results are available upon request.

**Competing interest:** The authors declare no competing financial interest.

# Supporting information

# The role of grain boundary scattering in reducing the thermal conductivity of polycrystalline XNiSn (X = Hf, Zr, Ti) half-Heusler alloys

*Matthias Schrade[1], Kristian Berland[1], Simen N. H. Eliassen[1,2], Matylda N. Guzik[1,3], Cristina Echevarria-Bonet[3], Magnus H. Sørby[3], Petra Jenuš[4], Bjørn C. Hauback[3], Raluca Tofan[1], Anette E. Gunnæs[1], Clas Persson[1], Ole Martin Løvvik[1,5], and Terje G. Finstad[1]*

[1] Centre for Materials Science and Nanotechnology, Department of Physics, University of Oslo, Gaustadalléen 21, NO-0349 Oslo, Norway

[2] Department of Materials Science and Engineering, Norwegian University of Science and Technology, Norway

[3] Physics Department, Institute for Energy Technology, NO-2007 Kjeller, Norway

[4] Jožef Stefan Institute, Department for Nanostructured Materials, Ljubljana, Slovenia

[5] SINTEF Materials and Chemistry, Forskningsveien 1, NO-0314 Oslo, Norway

## S.1. SR-PXD refinement results

1. <u>TiNiSn</u>

Phases identified from Rietveld analysis:

- **TiNiSn**: 95.8(4) % of total sample mass, crystallite size: d = 64 nm; microstrain : $\eta$ = 0.05(1)% ($R_{Bragg}$ = 5.08, $R_F$ = 2.52)
- **TiNi$_2$Sn**: 1.0(1) % of total sample mass ($R_{Bragg}$ = 16.9, $R_F$ = 10.5)
- **Sn** 3.2(1) % of total sample mass ($R_{Bragg}$ = 6.79, $R_F$ = 6.42)

$R_p$ = 6.93, $R_{wp}$ = 9.68, $R_{exp}$ = 0.01

## 2. ZrNiSn

Phases identified from Rietveld analysis:

- **ZrNiSn**: 100 % of total sample mass, crystallite size: d = 96 nm; microstrain : $\eta$ = 0.08(1)%

$R_p$ = 9.21, $R_{wp}$ =11.8, $R_{exp}$ = 0.01, $R_{Bragg}$ = 8.96, $R_F$ = 4.20

Though no other phase was included in the refinement there are peaks suggesting the presence of **Sn** but they account for not more than 1-2 % of total sample mass.

## 3. HfNiSn

Phases identified from Rietveld analysis:
- **HfNiSn**: 91.4(6) % of total sample mass, crystallite size: d = 49 nm; microstrain : $\eta$ = 0.008(3)% ($R_{Bragg}$ = 16.9, $R_F$ = 10.5)
- **HfNi$_2$Sn**: 0.04(2) % of total sample mass ($R_{Bragg}$ = 29.2, $R_F$ = 23.8)
- **Hf**: 7.9(3) % of total sample mass ($R_{Bragg}$ = 20.5, $R_F$ = 11.1)
- **Sn**: 0.7(3) % of total sample mass ($R_{Bragg}$ = 37.7, $R_F$ = 31.1)

$R_p$ = 5.24, $R_{wp}$ = 7.67, $R_{exp}$ = 0.02

## 4. (Ti,Zr)NiSn
Phases identified from Rietveld analysis:
5 Half-Heusler phases:

- **Ti$_{0.30}$Zr$_{0.70}$NiSn**: 32.8(6) % of total sample mass ($R_{Bragg}$ = 6.27, $R_F$ = 3.55)
- **Ti$_{0.82}$Zr$_{0.18}$NiSn**: 10.4(2) % of total sample mass ($R_{Bragg}$ = 10.2, $R_F$ = 7.61)
- **Ti$_{0.10}$Zr$_{0.90}$NiSn**: 13.2(4) % of total sample mass ($R_{Bragg}$ = 8.76, $R_F$ = 5.89)
- **Ti$_{0.58}$Zr$_{0.42}$NiSn**: 35.6(4) % of total sample mass ($R_{Bragg}$ = 6.72, $R_F$ = 4.04)
- **Ti$_{0.33}$Zr$_{0.67}$NiSn**: 8.1(1) % of total sample mass ($R_{Bragg}$ = 6.93, $R_F$ = 4.31)

And other minor impurities including **Sn**, not included in the analysis. Crystallite sizes *d* vary in the range of 20-100 nm among identified phases.

$R_p$ = 6.29, $R_{wp}$ = 7.80, $R_{exp}$ = 0.01

### 5. (Ti,Hf)NiSn

Phases identified from Rietveld analysis:

2 Half-Heusler phases:

- **Ti$_{0.31}$Hf$_{0.69}$NiSn**: 57.8(5) % of total sample mass, crystallite size: d = 56(3) nm; microstrain: η = 0.39(5) % (R$_{Bragg}$ = 6.79, R$_F$ = 5.01)
- **Ti$_{0.69}$Hf$_{0.31}$NiSn**: 42.2(6) % of total sample mass, crystallite size: d = 135(7) nm; microstrain: η = 0.21(4) % (R$_{Bragg}$ = 8.62, R$_F$ = 6.42) and other minor impurities including **Sn**, not included in the analysis.

R$_p$ = 5.37, R$_{wp}$ = 8.25, R$_{exp}$ = 0.02

### 6. (Zr,Hf)NiSn

Phases identified from Rietveld analysis:

- **Zr$_{0.62}$Hf$_{0.38}$NiSn**: 100 % of total sample mass, d = 79(1) nm; microstrain: η = 0.12(1)%

R$_p$ = 4.70, R$_{wp}$ = 5.51, R$_{exp}$ = 0.05, R$_{Bragg}$ = 9.02, R$_F$ = 5.18

### 7. (Ti, Zr,Hf)NiSn

Phases identified from Rietveld analysis:

- **Ti$_{0.40}$Hf$_{0.30}$Zr$_{0.30}$NiSn**: 37(1) % of total sample mass, crystallite size: d = 58(3) nm; microstrain: η = 0.24(4) % (R$_{Bragg}$ = 7.92, R$_F$ = 6.00)
- **Ti$_{0.60}$Hf$_{0.20}$Zr$_{0.20}$NiSn**: 63(2) % of total sample mass, crystallite size: d = 40(3) nm; microstrain: η = 0.18(3) % (R$_{Bragg}$ = 7.28, R$_F$ = 5.04)

And other minor impurity phases.

R$_p$ = 5.33, R$_{wp}$ = 7.61, R$_{exp}$ = 0.02

## S.2. TEM and SEM micrographs

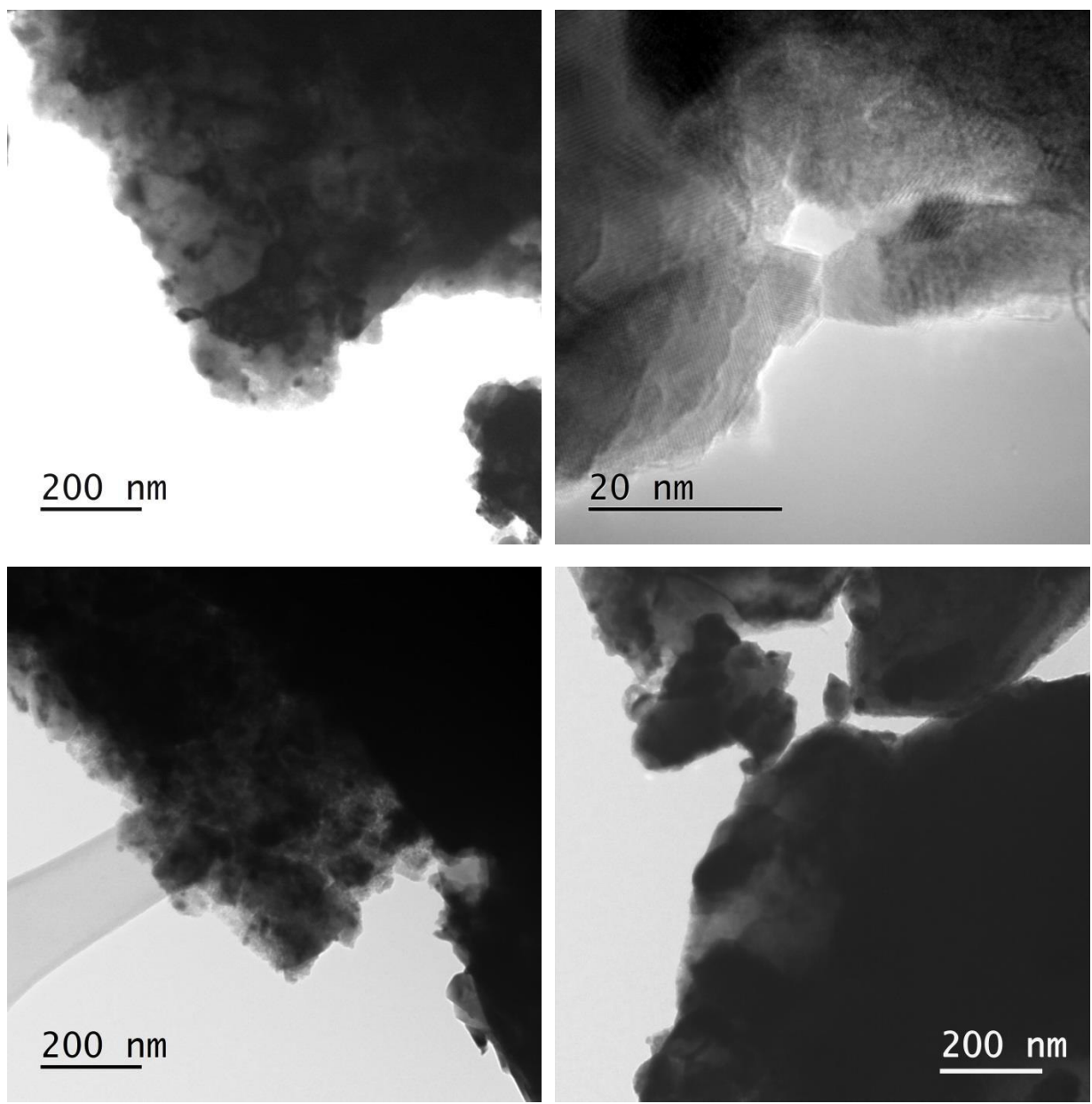

**Figure S.1.** TEM pictures of a ZrNiSn sample after SPS. Crystallites of varying size can be observed, in qualitative agreement with the volume-weighted average crystallite size as obtained by refinement of SR-PXD data.

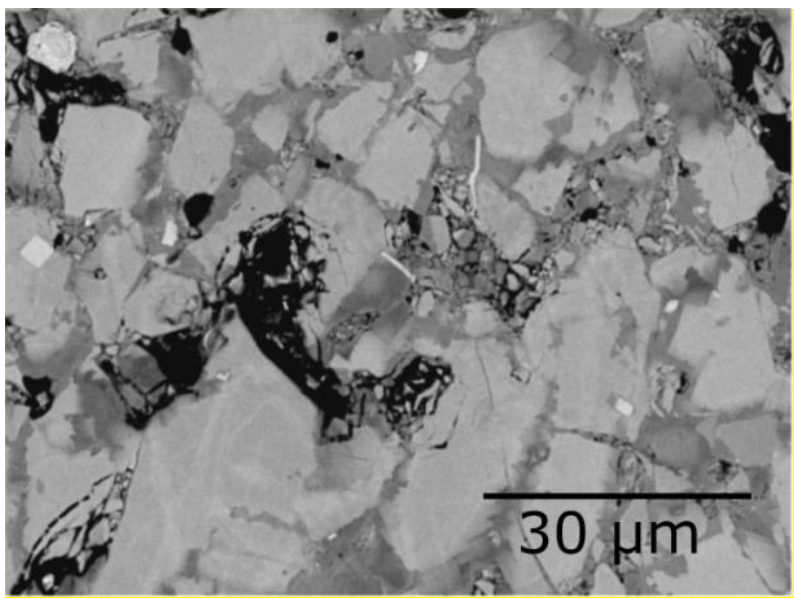

**Figure S.2.** SEM micrograph of the $Hf_{0.5}Ti_{0.5}NiSn$ sample. Phase contrast between two major compositions is visible. Bright areas correspond to a Hf-rich, dark areas to a Ti-rich phase. Composition as obtained by EDS agrees well with the results from PXD. The SEM images depict areas of homogeneous composition, with dimensions ranging from around 1 to 20 μm. These areas consist of several smaller crystallites below the resolution limit of the SEM used here, as illustrated by the TEM micrographs in Fig. S.1.

| Composition | $d_{Exp}$ [g/cm$^3$] | $d_{Theo}$ [g/cm$^3$] | Relative Density |
|---|---|---|---|
| HfNiSn | 10.32 | 10.51 | 98.2 % |
| ZrNiSn | 7.30 | 7.80 | 93.5 % |
| TiNiSn | 6.86 | 7.17 | 95.7 % |
| $Hf_{0.38}Zr_{0.62}NiSn$ | 8.29 | 8.85 | 93.7 % |
| $Zr_{0.5}Ti_{0.5}NiSn$ | 6.96 | 7.51 | 92.7 % |
| $Hf_{0.5}Ti0.5NiSn$ | 8.26 | 8.9 | 92.8 % |
| $Hf_{0.25}Zr_{0.25}Ti_{0.5}NiSn$ | 8.20 | 8.21 | 100 % |

**Table S.1.** Experimental gravimetric density $d_{Exp}$ compared to the theoretical density $d_{Theo}$ for the samples investigated here.

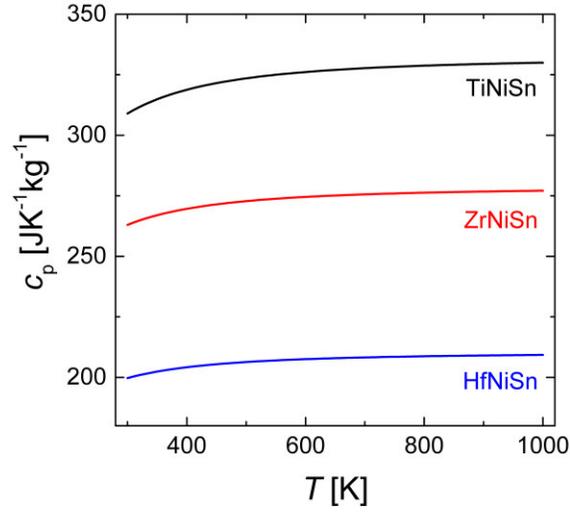

**Figure S.3.** Specific heat capacity $c_P$ as a function of temperature for the three unmixed compositions. Values for the mixed compositions are obtained by interpolating the unmixed compositions.

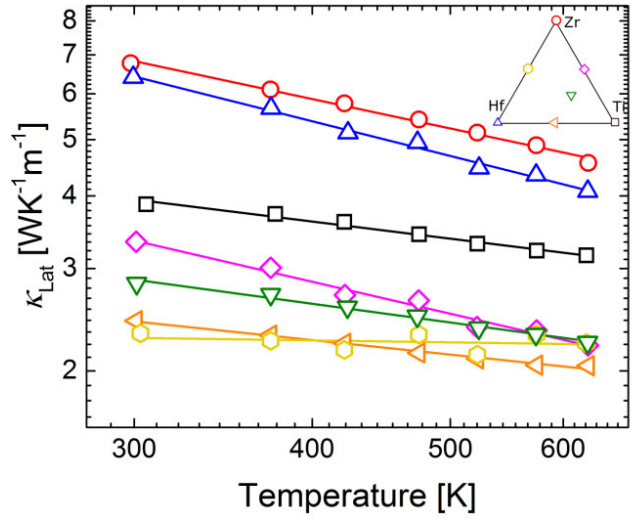

**Figure S.4.** Experimental $\kappa_{Lat}$ as shown in Fig. 5. Solid lines represent the best fit $\kappa_{Lat} \propto T^{-x}$, to obtain the temperature exponent $x$.